\documentclass{article}
\usepackage{etoolbox}
\usepackage{amsfonts}
\usepackage{amssymb}
\usepackage{pifont}
\newcommand{\cmark}{\ding{51}}%
\usepackage[T1]{fontenc} 
\usepackage[utf8]{inputenc} 
\usepackage{ismir_arxiv,amsmath,cite,url}
\usepackage{graphicx}
\usepackage{color}
\usepackage{booktabs}
\usepackage{lineno}
\usepackage{xspace}
\usepackage{bbm}
\usepackage{amsopn}
\usepackage{tabularx}
\usepackage{bm}

\newcommand{\vc}{\bm{c}}

\newcommand{\vq}{\bm{q}}

\newcommand{\vz}{\bm{z}}               

\newcommand{\my}{\bm{Y}}

    \newcommand{\Yc}{\mathcal{Y}}

\newcommand{\method}{\textsc{Jasco}\xspace}
\newcommand{\pmr}[1]{\scriptsize$\pm$#1}
\newcommand{\newpara}[1]{\vspace{0.36cm}\noindent \textbf{#1}}

\newtoggle{release}
\toggletrue{release}

\iftoggle{release}{
\newcommand{\adios}[1]{}
\newcommand{\alon}[1]{}
\newcommand{\ort}[1]{}
}
{
\newcommand{\adios}[1]{{\color{magenta}Y: #1}}
\newcommand{\alon}[1]{{\color{cyan}A: #1}}
\newcommand{\ort}[1]{{\color{blue}O: #1}}
}

\title{JOINT AUDIO AND SYMBOLIC CONDITIONING FOR \\ 
TEMPORALLY CONTROLLED TEXT-TO-MUSIC GENERATION}

\multauthor
{Or Tal$^{* 1,2}$\thanks{*Equal contribution} \hspace{1cm} Alon Ziv$^{* 1}$ \hspace{1cm} Itai Gat$^2$} { \bfseries{Felix Kreuk$^2$ \hspace{1cm} Yossi Adi$^{1,2}$ \hspace{1cm}}\\
    $^1$The Hebrew University of Jerusalem\\
    $^2$ Meta, FAIR Team\\
{\tt\small \{or.tal1, alon.ziv1\}@mail.huji.ac.il}
}

\def\authorname{O. Tal, A. Ziv, I. Gat, F. Kreuk, and Y. Adi}

\usepackage[bookmarks=false,pdfauthor={\authorname},pdfsubject={\papersubject},hidelinks]{hyperref}
\usepackage[capitalize]{cleveref}

\begin{document}

\maketitle
\begin{abstract}
We present \method, a temporally controlled text-to-music generation model utilizing both symbolic and audio-based conditions. \method can generate high-quality music samples conditioned on global text descriptions along with fine-grained local controls. \method is based on the Flow Matching modeling paradigm together with a novel conditioning method. This allows music generation controlled both locally (e.g., chords) and globally (text description). Specifically, we apply information bottleneck layers in conjunction with temporal blurring to extract relevant information with respect to specific controls. This allows the incorporation of both symbolic and audio-based conditions in the same text-to-music model. We experiment with various symbolic control signals (e.g., chords, melody), as well as with audio representations (e.g., separated drum tracks, full-mix). We evaluate \method considering both generation quality and condition adherence, using both objective metrics and human studies. Results suggest that \method is comparable to the evaluated baselines considering generation quality while allowing significantly better and more versatile controls over the generated music. Samples are available on our demo page \url{https://pages.cs.huji.ac.il/adiyoss-lab/JASCO}
\end{abstract}

\section{Introduction}
\label{sec:introduction}

\begin{figure}[ht!]
    \centering
    \includegraphics[width=\columnwidth]{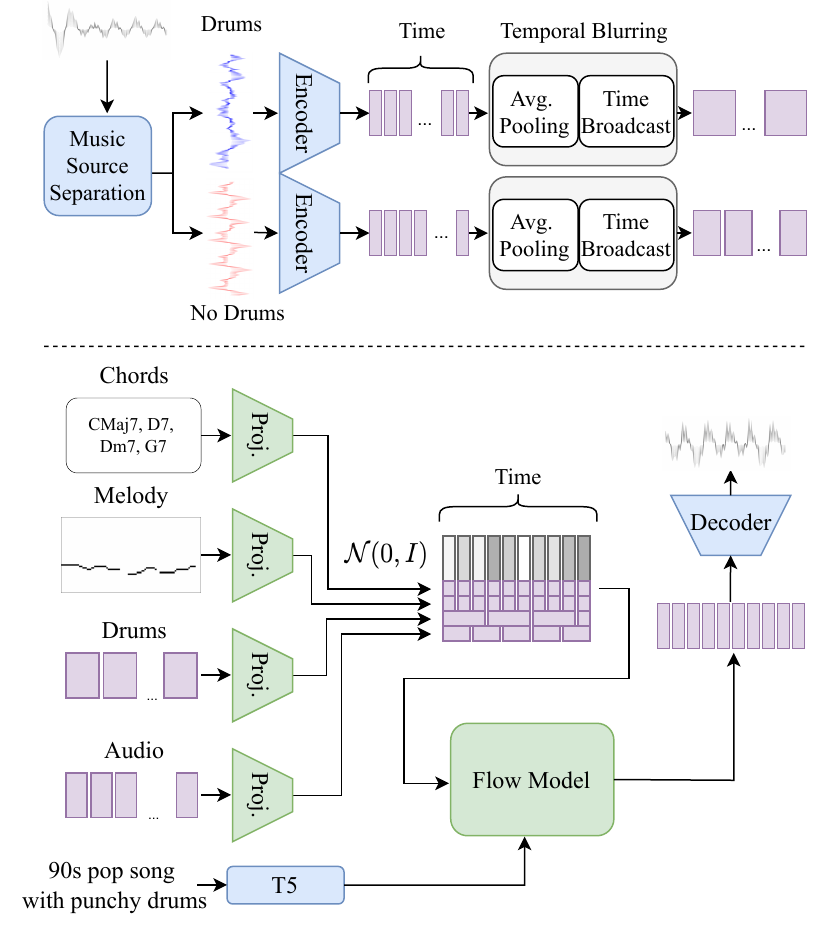}    
    \caption{Top figure presents the temporal blurring process, showcasing source separation, pooling and broadcasting. Bottom figure presents a high level presentation of \method. Conditions are first being projected to low dimensional representation and are concatenated over the channel dimensions. Green blocks have learnable parameters while blue block are frozen.\label{fig:arch}}
\end{figure}

Conditional music generation has shown a great improvement in recent years, specifically in the task of \textit{text-to-music} generation \cite{agostinelli2023musiclm, huang2023noise2music, copet2023simple, schneider2023mo, li2023jen, ziv2024masked}. Such advancements in music generation hold great potential to empower content creators, advertisers, and video game designers. Though presenting highly realistic music samples, most of the prior work is focused on global conditioning only. Such methods mainly consider textual descriptions or melody in the form of spectral features~\cite{copet2023simple}. However, when considering music production, global controls may not be enough. During the creative process, professional musicians often use chords, melodies, or audio prompts, at the local level, rather than global descriptions. As a result, current models may be limited in their relevancy for music creators.

More recently, several works study text-to-music generation using temporally aligned controls. The authors in~\cite{wu2023music} suggest adding symbolic beat and dynamics conditions on top of the previously explored melody conditioning. The authors in~\cite{novack2024ditto} further explore musical structure conditioning, such as A-part and B-part. Unlike these works, the proposed method provides local controls considering both symbolic representation and raw audio together with a global textual description. When considering music editing, the authors in~\cite{han2023instructme} propose leveraging chord progression to guide the generation process towards the harmony of the inputs signal. For that, the authors extract an internal representation from stemmed data using a pre-trained chord classification model. The proposed method is different as we focus on generating full musical pieces rather than editing a given one. Specifically, we allow symbolic chord progression conditioning during inference time.

In this work, we present \method, a locally controlled \underline{J}oint \underline{A}udio and \underline{S}ymbolic \underline{CO}nditioning text-to-music model. \method uses time-aligned controls, namely audio prompts, melodies and chord progressions, comprised of either symbolic signals or raw waveforms. We relieve the need for either studio quality stemmed data or supervised datasets by using off-the-shelf pre-trained models to automatically extract the relevant information. We use a source separation network~\cite{rouard2022hybrid} for drum extraction, an F0 saliency detector model~\cite{Bittner2017DeepSR} for melody extraction, and a chord progression extraction model~\cite{de2012improving} for harmonic conditioning.  We introduce a simple yet effective approach for audio conditioning using low-dimensional bottleneck projections, band pass filters, and temporal blurring. \method is based on the Flow-Matching~\cite{lipman2023flow} modeling paradigm. \Cref{fig:arch} provides a high level description of the proposed method. 

We compare \method to several baselines and provide a thorough analysis on the components composing \method. Results suggest \method provides comparable performance to the evaluated baselines considering generation quality while allowing significantly richer set of controls that can be used jointly or separately. 

\section{Background}
\label{sec:back}

\newpara{Audio Representation.} 
Modern audio generative models mostly operate on a latent representation of the audio, commonly obtained from a compression model~\cite{borsos2023audiolm, kreuk2022audiogen, yang2022diffsound}. Compression models such as \cite{defossez2022high} employ Residual Vector Quantization (RVQ) which results in several parallel streams. Each stream is comprised of discrete tokens originating from different learned codebooks. 

Specifically, the authors in \cite{defossez2022high} introduced EnCodec, a convolutional auto-encoder with a latent space quantized using RVQ~\cite{zeghidour2021soundstream}, and an adversarial reconstruction loss. Given a reference audio signal $x \in \mathbb{R}^{D \cdot f_s}$ with $D$ the audio duration and $f_s$ the sample rate, EnCodec first encodes it into a continuous latent tensor $\vz \in \mathbb{R}^{D \cdot f_r \times N_{\text{enc}}}$ with a frame rate $f_r \ll f_s$ and $N_{\text{enc}}=128$. Then, $\vz$  is quantized into $\vq \in \{1, \ldots, N\}^{D\cdot f_r \times K}$, with $K$ being the number of codebooks used in RVQ and $N$ being the codebook size. After quantization, we are left with $K$ discrete token sequences, each of length $T = D \cdot f_r$, representing the audio signal. In RVQ, each quantizer encodes the quantization error left by the previous quantizer, thus quantized values for different codebooks are in general dependent, where the first codebook holds most of the information. Finally, the quantized representation is decoded back to a time domain signal using the decoder network applied to the sum of the representations learned by the different codebooks. In \method, we use the continuous tensor $\vz$ as the latent representation, while leveraging the discrete representation $\vq$ for audio conditioning. 

\newpara{Flow Matching.} 
The Flow Matching modeling  paradigm \cite{lipman2023flow} was recently found to provide impressive results on image~\cite{lipman2023flow}, speech~\cite{le2023voicebox} and environmental sound generation~\cite{vyas2023audiobox}.  More specifically, Conditional Flow Matching (CFM) is a novel training technique for Continuous Normalizing Flow models \cite{chen2019neural}, that captures the continuous transformation paths of samples from a basic prior distribution, usually standard normal $\mathcal{N}(0,1)$, to their counterparts in a target data distribution, $\mathcal{S}$. The position on this path is denoted by a time parameter $t$, starting from the prior state at $t=0$ and ending at the data state at $t=1$. 

In this work, we focus on Optimal Transport (OT) paths as defined in \cite{lipman2023flow}. The model is trained to predict the vector field of the continuous latent audio variable $\vz$, given $t$ and a set of conditions $\my$. Formally, the model minimizes the regression loss
\begin{equation}
    \mathcal{L}_{\text{CFM}}(\theta ; \vz_0, \vz_1, t |  \my) = \lVert v_{\theta}(\vz, t | \my) - (\vz_1 - (1-\sigma_{\text{min}}) \cdot \vz_0)\rVert^2,
    \label{eq:orig_loss}
\end{equation}
where $\vz_0\sim \mathcal{N}(0,I)$ is a sampled noise, $\vz_1\sim \mathcal{S}$ is the latent representation of a data sample, and 
\begin{equation}
    \vz=(1-(1-\sigma_{\text{min}})\cdot t) \cdot \vz_0 + t \cdot \vz_1,
\end{equation} 
is an interpolation between the noise and the data sample. 
For numerical stability, we use a small value $\sigma_{\text{min}}=10^{-5}$ in both terms.
During inference we follow an iterative process, starting with the prior noise $\vz\leftarrow\vz_0 \sim \mathcal{N}(0,1)$ and with $t=0$. In each step, we translate the estimated vector field $v_\theta (\vz,t|\my)$ into an updated latent sequence $\vz$, and gradually converge toward the data distribution.
\section{Method}\label{sec:method}

Given a textual description, and a set of temporal conditions - such as melody, chord progression or drum recording, our goal is to produce high-quality samples that are musically aligned with the given controls, while complying to the arrangement description provided in the text. 

\method  tackles the aforementioned problem by a CFM model, operating on the continuous latent space of EnCodec. \method is conditioned on low-dimensional embeddings of  melody, chords and audio signals, together with a T5 embedding of the textual description. All local controls are concatenated to the model's input across the feature dimension, while text is being passed via cross attention. To diminish timbre-related information, \method further applies temporal blurring to the audio-based controls, as well as band-pass filtering. See~\Cref{fig:arch} for a visual description, and~\Cref{subsec:temp_cont} for detailed information.

\subsection{Temporal Controls}
\label{subsec:temp_cont}

\newpara{Symbolic.} We use Chordino~\footnote{\url{https://github.com/ohollo/chord-extractor}} chord progression model to extract an integer categorical chord label sequence, and a pretrained multi-F$0$ classifier~\cite{Bittner2017DeepSR} to obtain melody scores per time step. We resample all features to match EnCodec's frame rate using 'nearest' interpolation for chords and 'linear' interpolation for melody. For Chords, we use a learned embedding table to map the raw integer sequence, denoted as $\vc_{\text{crd}}$, to its corresponding condition matrix $\in\mathbb{R}^{T\times d_{\text{crd}}}$.
For Melody, we zero out values with a score lower than a pre-defined threshold ($0.5$). Then, we select the maximal non-zero score per time step from the remaining values, and set it to $1$ while setting the rest to $0$. This yields a binary matrix  $\vc_{\text{mld}} \in \{0,1\}^{D\cdot f_r^{\text{mld}} \times N_{\text{mld}}}$. Finally, we linearly project the binary matrix and obtain the melody condition representation $\in\mathbb{R}^{T\times d_{\text{mld}}}$.
We use $N_{\text{mld}}=53$ (corresponding to G$2$-B$7$ notes), and $d_{\text{crd}}=d_{\text{mld}}=16$.

\newpara{Audio.} We consider general audio and separated drum stems. We use a pretrained source separation model~\cite{defossez2021hybrid}, to extract the drum stem from a source audio. We pass the waveform through EnCodec to obtain the corresponding quantized discrete representation $\vq$. We then convert the first token stream back to its continuous latent representation, using EnCodec's first codebook while discarding all other streams, yielding $\vc_{\text{aud}},\vc_{\text{drm}}\in\mathbb{R}^{T \times N_{\text{enc}}}$. Following that, we apply temporal blurring to the reconstructed latent sequence. First, we perform average pooling using non-overlapping windows along the temporal axis. Then, we broadcast the signal back to its original temporal dimension. Finally, we linearly project the blurred condition to a low dimensional feature space and obtain the final condition matrix. For the general audio condition, we use a window size of $5$ and output dimension of $1$, while for drums we use a window size of $3$ and output dimension of $2$. 

\newpara{Inpainting and Outpainting.} Following  prior work~\cite{li2023jen}, we add in/out-painting as an additional condition to the model. We randomly choose between inpainting/outpainting, and mask a random segment of  $40$-$90$\% from the reference waveform. Then, we use the raw EnCodec latent representation of the masked waveform $\vc_{\text{iop}}\in\mathbb{R}^{T \times N_{\text{enc}}}$ as the condition, with no learned projection.
\vspace{-0.2cm}
\subsection{Model and Optimization}
Similarly to prior work~\cite{vyas2023audiobox}, our CFM model consists of a Transformer, with U-Net-like residual connections. We replace the standard residual addition with channel-wise concatenation followed by a linear projection. We use learned convolutional positional encoding \cite{baevski2020wav2vec} as well as symmetric bi-directional ALiBi self-attention biases~\cite{press2022train}. We use a model scale of $330$M parameters, with $24$ Transformer layers, $16$ attention heads, embedding dimensionality of $1024$ and a feed-forward dimension of $4096$. 

We train our model using the $\mathcal{L}_{\text{CFM}}$ objective as defined in \Cref{sec:back}. For a batch of samples, we further experiment with non-uniform loss weighting as function of $t$, and find the following formulation to produce the best overall sample quality:
\begin{equation} 
    \mathcal{L}_{\text{WeightedCFM}} = \sum_{\substack{t\sim \mathcal{U}(0,1)\\
                  \vz_0\sim \mathcal{N}(0,1)\\
                  \vz_1 \sim \mathcal{S}}}
        (1 + t) \cdot \mathcal{L}_{\text{CFM}}(\theta ; \vz_0, \vz_1, t | \my),
        \label{eq:loss}
\end{equation} 
where $\my = \{\vc_{\text{crd}}, \vc_{\text{mld}}, \vc_{\text{aud}}, \vc_{\text{drm}}, \vc_{\text{iop}}\}$. We provide an ablation study for this scheme in \Cref{sec:results}.

\subsection{Inference}

During inference, we use \textit{dopri5} \cite{dormand1980family}, an off-the-shelf numerical ODE solver, to iteratively solve for $\vz$ given the estimated vector field $v_{\theta}$. Specifically, at each iteration the solver determines the increment to the time parameter $t$, resulting in a dynamic scheduling for the inference process. The process halts when an acceptance criterion is met, defined by an error approximation of the solver and a tolerance parameter provided by the user.

\newpara{Multi-Source Classifier Free Guidance.}
We employ classifier-free guidance (CFG)~\cite{ho2022classifierfree} for the conditional vector field estimation $v_{\theta}(\vz,t|\Yc)$. Since our set of conditioning signals combines both global and local concepts, we further experiment with multi source CFG. While prior work \cite{parker2024stemgen} suggest a separate evaluation for each condition, we evaluate the model considering all and partial conditions. During each inference step, we obtain an estimated vector field for each set of conditions $\Yc \in \{\{\textrm{local}\}, \{\textrm{text}\}, \{\textrm{local, text}\}\}$. The resulting CFG formulation then follows:
\begin{equation}
\text{CFG}(v_\theta, \vz, t)\text{=}(1 - \sum_{c \in \Yc}\alpha_{c})v_\theta(\vz, t)  + \sum_{c \in \Yc}\alpha_{c}v_\theta(\vz, t | c).
\end{equation}

When following the standard CFG setup ($\alpha_{\textrm{text}}=\alpha_{\textrm{local}}=0$), we observe that the model adheres to the temporal condition while ignoring instrumentation information provided in the text prompt. To increase text influence on guidance, we set a positive weight to the text-only term $\alpha_{\textrm{text}}>0$. We found that $\alpha_{\textrm{text}}=0.5, \alpha_{\textrm{local}}=0, \alpha_{\textrm{local,text}}=1.5$ offer a good trade-off between audio quality, text alignment and temporal controls adherence. 
\section{Experimental Setup}\label{sec:exp_setup}
\newpara{Implementation Details.} We follow the same experimental setup as in \cite{copet2023simple,ziv2024masked}, and use a training dataset consisting of $20$K hours of licensed music from the Shutterstock \footnote{\url{shutterstock.com/music}} and Pond$5$ \footnote{\url{pond5.com}} data collections with $25$K and $365$K instrument-only music tracks, respectively. We additionally include a set of proprietary data consisting of $10$K high-quality tracks. All datasets are sampled at $32$kHz, paired with textual descriptions. We present results on the MusicCaps benchmark \cite{agostinelli2023musiclm}, comprising $5.5$K $10$-second samples together with an in-domain test set of $528$ tracks. 

We use the official EnCodec model provided by \cite{copet2023simple, kreuk2022audiogen}, with a frame rate of $50$ Hz, and $4$ codebooks, each with a size of $2048$. For text representation we use a pretrained T5 model~\cite{raffel2020t5}. For melody extraction we use the pretrained deep salience multi-F0 detector \footnote{\url{github.com/rabitt/ismir2017-deepsalience}}, for chords extraction we use Chordino, while for drum track extraction we use the Hybrid Demucs model~\cite{rouard2022hybrid}.

All single condition models were trained with $40$\% condition dropout, and in the multi-condition experiments we train the models with $20\%$ condition dropout for all conditions. In the remaining $80$\% we set $50$\% dropout for each of the conditions independently excluding the in/out-painting, for which we set $70$\% dropout.

We experiment with multi-source CFG coefficients in $(\alpha_{\textrm{text}}, \alpha_{\textrm{local}}, \alpha_{\textrm{text,local}})\in \{0.0, 0.5\}\times\{0.0, -0.5\}\times\{1.5, 2.0\}$ and report the best overall configuration. All models were trained for $500$k steps over audio segments of $10$ seconds, with a batch size of $336$. We use Adam \cite{kingma2017adam} optimizer with linear learning rate warm-up up to a peak of $10^{-4}$ during the first $5$k steps, followed by a linear decay, and a gradient clipping with a norm threshold of $0.2$.

\subsection{Evaluation Metrics}
\label{subsec:eval_metrics}
We perform a thorough empirical evaluation, using both objective metrics and human studies. We evaluate \method on several temporal alignment aspects, namely harmonic matching, rhythmic alignment and melody preservation. Additionally, we measure audio quality and text adherence. 

\newpara{Objective Evaluations.} 
We evaluate our method with widely used metrics, namely Fréchet Audio Distance (FAD), Kullback-Leiber Divergence (KL) and CLAP score (CLAP), as well as more specific metrics designed to quantify the adherence of our suggested controls. We report FAD~\cite{kilgour2018fr} using the official tensorflow implementation where a low FAD score indicates that the generated audio is associated with higher quality. Following \cite{kreuk2022audiogen,copet2023simple}, we use an audio classifier~\cite{koutini2021efficient} to compute the KL-divergence over the probabilities of the labels between the original and the generated music. The generated music is expected to share similar concepts with the reference music when the KL is low. Last, CLAP score~\cite{wu2023large, huang2023make} is computed between the track description and the generated audio, measuring audio-text alignment. We use the official pretrained CLAP model\footnote{\url{github.com/LAION-AI/CLAP}}. To evaluate melody compatibility, similar to~\cite{copet2023simple} we use a cosine similarity metric on either a simple quantized chroma representation, or multi-octave melody representation obtained from a pretrained multi-F0 classifier~\cite{Bittner2017DeepSR}. For beat adherence, as in~\cite{wu2023music} we evaluate the onset F1 score using \textit{mir eval}\footnote{\url{github.com/craffel/mir_evaluators}} considering a $50$ms tolerance margin around classified onsets in the reference signal. 
Lastly, to evaluate chord progression, we use the Chordino model to extract the chord progression from both the reference and the generated signals and compute the intersection over union (IOU) score between the two.

\begin{table}[t!]  
    \centering
    \resizebox{1.0\columnwidth}{!}{
    \begin{tabular}{l|ccccc}  
    \addlinespace
    \toprule
    \textbf{Model} & \textbf{FAD}$\downarrow$ & \textbf{CLAP}$\uparrow$ & \textbf{Mel Sim.}$\uparrow$ & \textbf{Mel Acc.}$\uparrow$\\
    \midrule
    MusicGen                   &  5.90   &  0.29   &  0.61 &  44.0  \\
    MusicControlNet            &  10.81  &  0.22   &  -    &  47.1  \\
    \method                    &  6.05   &  0.26   &  0.67 &  49.1  \\
    \bottomrule
    \end{tabular}}
    \caption{Melody conditioning evaluation over MusicCaps. We evaluated MusicGen with $300$M parameters.\label{tab:chroma comparison}}
\end{table}

\begin{table*}[t!]
  \centering
  \resizebox{\textwidth}{!}{\begin{tabular}{cccc|cccc|ccc}
  \addlinespace
    \toprule
    \multicolumn{4}{c|}{\textit{Local Controls}} & \multicolumn{7}{c}{\textit{Objective metrics (MusicCaps / Internal dataset)}} \\
    \bf Aud & \bf Drm & \bf Crd & \bf Mld   &  \bf Mld (clf) sim. $\uparrow$ & \bf Mld sim. $\uparrow$ & \bf Onset F1 $\uparrow$ & \bf Crd IOU $\uparrow$  & \bf FAD $\downarrow$ & \bf KL $\downarrow$  & \bf CLAP $\uparrow$ \\
    \midrule
    -       & -       & -       & -        & 0.13 / 0.13 &  0.09 / 0.09  &  0.34 / 0.41  &  0.09 / 0.07 & 6.04 / 0.90  &  1.46 / 0.70 &  0.27 / 0.36  \\
    \midrule
     \cmark  & -       & -       & -       & 0.33 / 0.34 &  0.38 / 0.47  &  0.62 / 0.81  &  0.23 / 0.27 & 4.47 / 0.86  &  0.92 / 0.81 &  0.30 / 0.31  \\  
     no drm  & -       & -       & -       & 0.21 / 0.22 &  0.38 / 0.31  &  0.62 / 0.58  &  0.23 / 0.18 & 5.68 / 0.92  &  1.79 / 0.75 &  0.19 / 0.33  \\  
     -       & \cmark  & -       & -       & 0.13 / 0.13 &  0.09 / 0.10  &  0.62 / 0.73  &  0.09 / 0.08 & 5.85 / 0.94  &  1.68 / 0.78 &  0.23 / 0.35  \\  
     -       & BPF     & -       & -       & 0.13 / 0.13 &  0.10 / 0.10  &  0.45 / 0.74  &  0.10 / 0.07 & 6.31 / 1.61  &  1.52 / 0.65 &  0.26 / 0.37  \\  
     -       & -       & \cmark  & -       & 0.21 / 0.25 &  0.22 / 0.29  &  0.24 / 0.13  &  0.59 / 0.61 & 7.23 / 0.95  &  1.16 / 0.68 &  0.28 / 0.36  \\  
     -       & -       & -       & \cmark  & 0.67 / 0.64 &  0.41  / 0.35 &  0.37 / 0.57  &  0.31 / 0.27 & 6.96 / 1.05  &  1.32 / 0.63 &  0.27 / 0.35  \\  
\midrule
      -      & BPF     &\cmark   & \cmark  & 0.68 / 0.69 &  0.44 / 0.46  &  0.63 / 0.66  &  0.50 / 0.53 & 6.42 / 1.15  &  1.22 / 0.50 &  0.28 / 0.37 \\
     no drm  & BPF     &\cmark   & \cmark  & 0.71 / 0.68 &  0.50 / 0.55  &  0.54 / 0.75  &  0.51 / 0.55 & 4.78 / 0.80  &  0.93 / 0.41 &  0.30 / 0.37 \\
    \bottomrule
  \end{tabular}}
  \caption{Objective local controls experiment, observing all suggested controls w.r.t a zero hypothesis (no local controls).}
  \label{tab:uni_cond}
\end{table*}

\newpara{Human Study.} We request raters to evaluate three aspects of given audio samples: (i) overall quality; (ii) similarity to text description; and (iii) adherence to either melody or rhythmic pattern from a reference recording. Raters were instructed to rate the recordings on a scale between $0$-$100$ where higher is better. Raters were recruited using the Amazon Mechanical Turk platform. We evaluate randomly sampled files, where each sample was evaluated by at least $5$ raters. We use the CrowdMOS package~\cite{ribeiro2011crowdmos} to filter noisy annotations and outliers. We remove annotators who did not listen to the full recordings, annotators who rate the reference recordings less than $90$, and the rest of the recommended recipes from~\cite{ribeiro2011crowdmos}. Similarly to~\cite{copet2023simple}, for a fair comparison, all samples are normalized at -$14$dB LUFS~\cite{sugimoto2022loudness}.

\vspace{-0.2cm}
\section{Results}
\label{sec:results}
\vspace{-0.1cm}

\newpara{Melody Conditioning.} We start by evaluating the proposed method considering melody conditioning. We compare \method to MusicGen~\cite{copet2023simple} and MusicControlNet~\cite{wu2023music}. For a fair comparison, we train MusicGen ($300$M) on $10$ second music segments using Audiocraft\footnote{\url{https://github.com/facebookresearch/audiocraft/blob/main/docs/MUSICGEN.md}}repository, considering text and melody conditions. For comparison compatibility with~\cite{wu2023music} we compute melody accuracy score on both \method and MusicGen. We experiment with melody conditioning using the commonly used $12$-bins chroma representation which is octave invariant. Results are presented in~\Cref{tab:chroma comparison}.

Results suggest that \method surpasses the evaluated baselines w.r.t melody adherence. When considering melody accuracy, \method provides better alignment to the conditioning melody. Notice, we hypothesize this is due to the conditioning method: both MusicGen and MusicControlNet inject conditions as an additive bias (i.e., cross-attention and zero-convolutions), this is in contrary to \method which follows the concatenation approach for melody conditioning (see \Cref{sec:analysis} for additional experiments). 

\newpara{Local Controls.} Next, we perform a thorough evaluation of \method for each of the suggested temporal controls, namely Chords, Melody, Audio, and Drums. We train a single-condition variant for each observed condition-type as well as two multi-condition models. Under the multi-condition setup, we train models with Drums tracks passed through a Band-Pass-Filter (BPF) over $200$-$800$ Hz frequency range, and Audio condition excluding drums. This was found to better disentangle Drums and Audio conditions in preliminary experiments, and allows users to provide different drum beats than the one presented in the Audio. When applying Audio/Drums conditions, we evaluate Melody, Onset F1, and Chord IoU using the reference audio as a condition, while for the computation FAD, KL, and CLAP scores we use a randomly selected audio from the test set as a condition. 

As there are no open-source relevant baselines available, we compare the proposed method against a text-only condition model. We perform experiments using both the open source MusicCaps dataset, and an internal proprietary dataset, highlighting our model performance on diverse, high quality recordings. Table~\ref{tab:uni_cond} summarizes the results.

Results depict a systematic improvement considering local control adherence. For instance, chords conditioning on both datasets show apparent improvement in Chords IOU metric, improving from $0.09 / 0.07$ to $0.59 / 0.61$. In addition, in spite of being evaluated with randomly selected audio conditions, FAD, KL, CLAP scores mostly remain comparable w.r.t to the baseline. This highlight \method's disentangling property as local controls metrics improve while text adherence and audio quality metrics stay roughly the same. 

The lower section of the table presents multi-control setup results. This section draws a similar trend to the single control setups, allowing for multiple controls while maintaining FAD, KL, CLAP scores. This highlights \method's ability to incorporate multiple controls simultaneously with no significant penalty to quality and text alignment. 

\begin{table}[t!]  
  \centering
  \resizebox{\columnwidth}{!}{
  \begin{tabular}{l|l|cccc}
  \addlinespace
    \toprule
    \textbf{Model} & \textbf{Cond.} & \textbf{Q} & \textbf{T} & \textbf{M} & \textbf{D} \\
    \midrule
    Reference & - & 92.7\pmr{0.66}& 93.7\pmr{0.8} & 96.3\pmr{0.6} & 97.1\pmr{0.6}\\
    \midrule
    MusicGen        &  T  & 84.4\pmr{0.8} & 84.5\pmr{0.9} & 81.5\pmr{1.3} & 82.1\pmr{1.0}\\
    \method         &  T  & 83.3\pmr{0.7} & 80.3\pmr{1.3} & 79.7\pmr{1.5} & 81.5\pmr{1.1}\\
    \midrule
    MusicGen        & T \& M & 84.7\pmr{0.7} & 82.5\pmr{1.1} & 83.6\pmr{1.1} & 82.7\pmr{0.9}\\
    \method         & T \& M & 84.1\pmr{0.7} & 81.2\pmr{1.2} & 89.3\pmr{0.7} & 80.6\pmr{1.2}\\
    \midrule
    \method         & T \& D  & 85.5\pmr{0.8} & 84.1\pmr{1.1} & 81.9\pmr{1.4} & 89.5\pmr{0.7}\\
    \bottomrule
  \end{tabular}}
  \caption{Human evaluation results. Observing general quality (Q), text match (T) melody match (M) and drums match (D). Evaluated on a 0-100 scale (higher is better).
  \label{tab:mos_unicond}}
\end{table}

\newpara{Human Study.} Lastly, we perform a human study in order to validate both quality and text alignment as well as local control adherence. We evaluate \method vs MusicGen considering: (i) text only; and (ii) both text and melody. We additionally, provide results of the proposed method with text and drums conditions. Results seen on~\Cref{tab:mos_unicond}, indicate that \method achieve similar generation quality as MusicGen across all setups. As of text relevancy, MusicGen reaches superior performance to the proposed method, however, when considering melody conditioning, \method reaches significantly better scores. Lastly, when conditioned on drums, \method provides the best rhythmic pattern similarity scores. This highlights \method's ability to provide better controls over the generated music without sacrificing quality and text alignment. Interestingly, after including melody or drums conditions, as expected, the relevant metrics are improving (i.e., melodic and rhythmic similarity) while the quality and text adherence remain comparable to the unconditioned model. 

\vspace{-0.2cm}
\section{Analysis}
\label{sec:analysis}
\vspace{-0.1cm}

\newpara{Condition Injection Method.} We compare the proposed method to two widely used condition injection methods proposed in prior work. Specifically, we perform a controlled experiment in which we evaluate cross-attention as used in MusicGen, and zero-convolution as used in MusicControlNet, considering the same training configuration. 

Results shown in~\Cref{tab:abl_cond} suggest that the temporal adherence using the concatenation method performs the best overall. This can be seen in both higher Chord IoU, as well as better FAD and KL, where CLAP was $0.36$ for all methods. Additional advantages for the concatenation method is the ability to train from scratch (as opposed to zero-convolutions, in which we start from a pretrained model) without a significant increase in the number of trainable parameters.

\begin{table}[t!]  
  \centering
  \resizebox{0.95\columnwidth}{!}{
  \begin{tabular}{l|ccc}
  \addlinespace
    \toprule
    \bf Conditioning & \bf Chord IOU $\uparrow$ & \bf FAD $\downarrow$ & \bf KL $\downarrow$ \\
    \midrule
    Concat              &  0.6   & 1.19 & 0.71 \\
    Cross Attn.         &  0.59  & 1.61	& 0.73  \\
    Zero Conv           &  0.26  & 1.64 & 0.74  \\
    \bottomrule
  \end{tabular}}
  \caption{Ablation for conditioning method. evaluated on internal dataset. All models started from a text-to-music pretrained checkpoint and trained for 500K steps.\label{tab:abl_cond}}
\end{table}

\begin{figure}[t!]
    \centering
    \includegraphics[width=1\columnwidth]{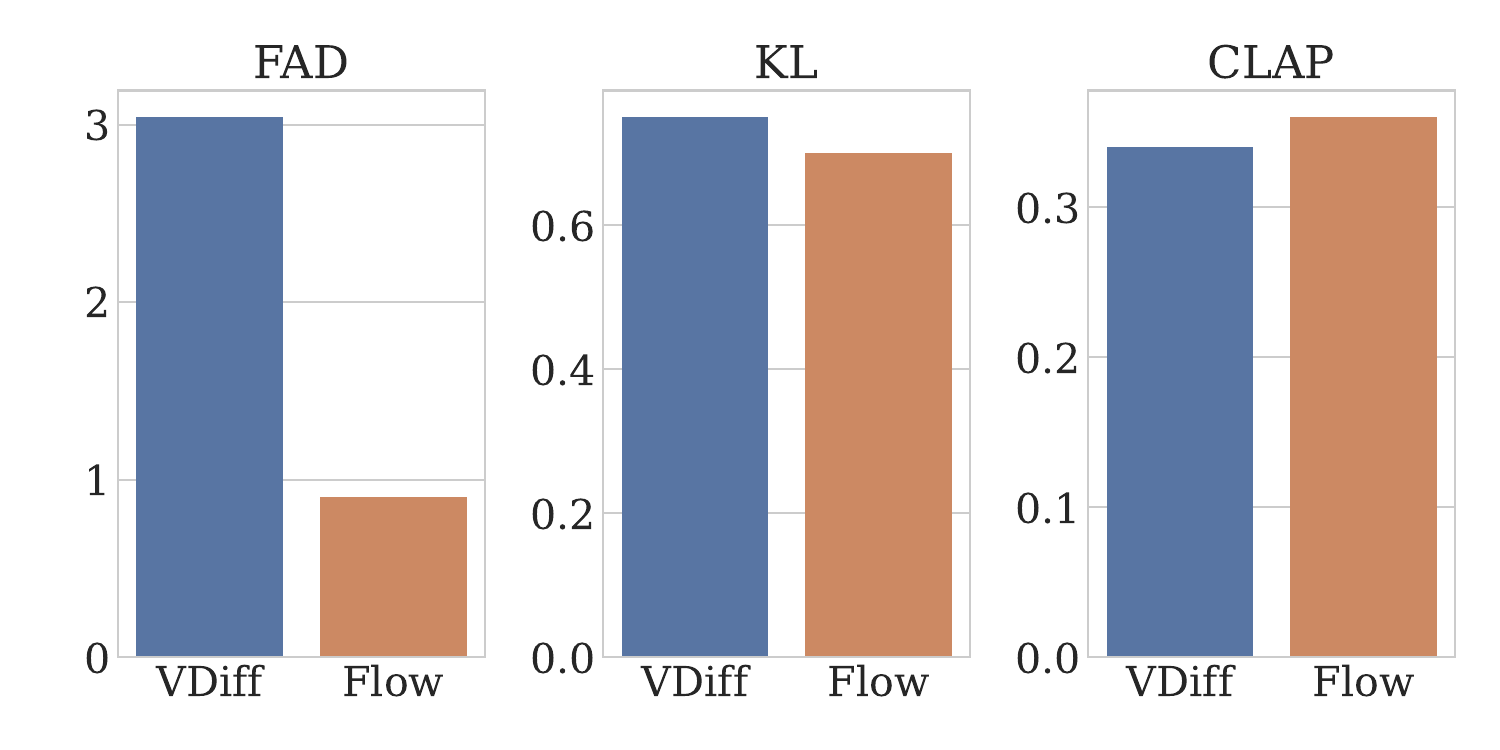}
    \caption{Comparison of v-Diffusion vs Flow Matching. We report FAD, KL, and CLAP on the internal dataset.} 
    \label{fig:vdif_vs_flow}
\end{figure}

\newpara{Flow vs. Diffusion.} Most of prior work on music generation is mainly based on Diffusion models~\cite{forsgrenriffusion, huang2023noise2music, schneider2023mo, li2023jen}. In this experiment we evaluate, under controlled settings, both Diffusion (v-Diffusion) and Flow Matching modeling approaches for music generation. We report FAD, KL, and CLAP scores. Results are depicted in \Cref{fig:vdif_vs_flow}. As can be seen, the Flow Matching approach is superior across all metrics, with the biggest gap observed in FAD. 

\newpara{The Effect of Weighted Loss.} Finally, we evaluate the effect of the proposed modification to the loss function as presented in~\Cref{eq:loss}. We compare the proposed objective function against the loss as describe in~\Cref{eq:orig_loss}, considering FAD, KL, and CLAP scores in~\Cref{tab:abl_loss}. Results suggest the new objective function modification  improves the generation quality. It provides significantly better FAD while having comparable KL and CLAP scores.

\section{Related Work}
\label{sec:related}

\newpara{Flow Matching for Audio Generation.} Flow Matching~\cite{lipman2023flow} was recently studied for speech generation. A notable work in this context presented VoiceBox~\cite{le2023voicebox}, a Flow Matching model, operating on spectrograms, for text-guided multilingual speech generation. More recently, AudioBox ~\cite{vyas2023audiobox} was presented, in which self-supervised infilling objectives were leveraged to improve the generalization capabilities of VoiceBox. Similar to our model, AudioBox operates on the continuous latent representations of EnCodec \cite{defossez2022high}. Though the scope of audio modalities was extended in AudioBox to both speech and environmental sounds, applying a Flow Matching approach for music generation remained less explored.

\newpara{Temporally Controlled Music Generation.} Recent work offered several forms of temporally restrictive controls for music generation. Melody conditioned text-to-music was studied in MusicLM \cite{agostinelli2023musiclm}, in which a melody embedding was trained using a dedicated dataset consists of multiple cover versions of musical tracks paired with aligned singing and humming performances. In MusicGen \cite{copet2023simple} and Music ControlNet \cite{wu2023music}, the need for supervised data was relieved, and instead an unsupervised melody extraction was performed using the argmax note of the audio chromagram. Audio-to-audio setups were studied for drum generation conditioned on drumless track \cite{wu2022jukedrummer}, accompaniment generation given singing voice \cite{donahue2023singsong}, and single instrument generation given partial mix \cite{parker2024stemgen} \cite{han2023instructme}. Recently, generation conditioned on multiple symbolic controls was studied in Music ControlNet \cite{wu2023music}, a spectrogram diffusion text-to-music model, fine-tuned using the ControlNet scheme \cite{zhang2023adding}, to generation with melody, beat and dynamics controls. In DITTO \cite{novack2024ditto}, inference time optimization was explored, for tiding a text-to-music diffusion model to perform several tasks including inpainting, outpainting, loop generation, melody and dynamics conditioned generation, as well as conditioning on musical structures.  In \cite{levy2023controllable}, classifier guidance was used to perform music inpainting, outpainting and style transfer given a pretrained unconditional latent diffusion model. Inpainting was further explored in \cite {li2023jen}, \cite{garcia2023vampnet}, and \cite{lin2024arrange}. Style transfer was explored also in \cite{mor2018universal} and \cite{han2023instructme}. 

\begin{table}[t!]  
  \centering
  \resizebox{0.95\columnwidth}{!}{
  \begin{tabular}{l|ccc}
  \addlinespace
    \toprule
    \textbf{Weighted loss schedule} & \textbf{FAD}$\downarrow$ & \textbf{KL}$\downarrow$ & \textbf{CLAP}$\uparrow$ \\
    \midrule
    $w(t) = 1$              &   1.73  &  0.71  &   0.38   \\
    $w(t) = 1 + t$          &   0.99  &  0.73  &   0.37   \\
    \bottomrule
  \end{tabular}}
  \caption{Ablation for loss weighting method. Evaluated on internal dataset. All models were trained for 500K steps.\label{tab:abl_loss}}
\end{table}
\section{Discussion}
\label{sec:con}

In this work we present \method, a temporally controlled text-to-music generation model, supporting both audio and symbolic conditioning. \method is based on the Flow Matching modeling paradigm operating over a dense music latent representation. Through extensive experimentation we empirically show \method generates high-fidelity samples that can be conditioned on global textual description together with harmony, melody, rhythmic patterns, and overall musical style. 
Results suggest \method provides comparable generation quality to the evaluated baselines while allowing significantly better control over generation.

\newpara{Limitations.} The main limitations of the proposed approach are: (i) Similarly to previous diffusion-based text-to-music models, the length of the generated samples is relatively short ($\sim10$ seconds) compared to the auto-regressive alternative. Although this can be extrapolated with overlaps, it may limit the capability of the model in capturing global structure in the generated music; (ii) although generating the whole sequence at once, generation time is slower than auto-regressive alternatives, while not supporting streaming capabilities. 

\newpara{Future work.} For future work we intend to support additional controls, such as music dynamics, musical structure, etc. together with editing options, e.g., add or replace specific instrument in a given recording. We believe such a research direction, and specifically the proposed approach, holds great potential in empowering musicians, creators, and producers which require richer set of controls during their creative process. 

\section{Ethical statement}
The use of large-scale generative models raises several ethical concerns. To mitigate at some of them, we first made sure all the data used for training our models was obtained legally through an agreement with ShutterStock. Another issue is the potential lack of diversity in the dataset, which predominantly consists of western-style music. However, we believe that the proposed method is not tied to any specific genera and can help expand the scope of applications to new datasets.

Moreover, generative models could potentially create an unbalanced competitive environment for artists, a problem that is yet to be solved. We are firm believers in the power of open research to provide all participants with equal opportunities to access these models. By introducing more sophisticated controls, like chords and rhythmic patterns as suggested in this work, we aspire to make these models beneficial for both amateurs and professional musicians.

\bibliography{refs}

\end{document}